\documentclass[aps,reprint,floatfix,twocolumn,amsmath,amssymb,superscriptaddress]{revtex4-1}

\usepackage{comment}
\usepackage{color}
\usepackage{url}
\usepackage{graphicx}% Include figure files
\usepackage{dcolumn}% Align table columns on decimal point
\usepackage{bm}% bold math
\usepackage[normalem]{ulem}

\usepackage[T1]{fontenc}
\usepackage[utf8]{inputenc}
\usepackage[english]{babel}

\newcommand{\bge}{\begin{equation}}
\newcommand{\ene}{\end{equation}}
\newcommand{\bgea}{\begin{eqnarray}}
\newcommand{\enea}{\end{eqnarray}}

\newcommand{\bfbb}{\mathbf{B}}

\newcommand{\bfm}{\mathbf{m}}

\newcommand{\bfbf}{\mathbf{B}^{\mathrm{fl}}}

\begin{document}

%\preprint{APS/123-QED}

\title{Spin wave excitations of magnetic metalorganic materials}

\author{Johan Hellsvik}
\email{hellsvik@kth.se}
\affiliation{Nordita, KTH Royal Institute of Technology and Stockholm University, Roslagstullsbacken 23, SE-106 91 Stockholm, Sweden}

\author{Roberto D\'iaz P\'erez}
\affiliation{Nordita, KTH Royal Institute of Technology and Stockholm University, Roslagstullsbacken 23, SE-106 91 Stockholm, Sweden}

\author{R. Matthias Geilhufe}
\affiliation{Nordita, KTH Royal Institute of Technology and Stockholm University, Roslagstullsbacken 23, SE-106 91 Stockholm, Sweden}

\author{Martin M{\aa}nsson}
\affiliation{Department of Applied Physics, KTH Royal Institute of Technology, Electrum 229, SE-164 40 Kista, Sweden}

\author{Alexander V. Balatsky}
\affiliation{Nordita, KTH Royal Institute of Technology and Stockholm University, Roslagstullsbacken 23, SE-106 91 Stockholm, Sweden}

\date{\today}% It is always \today, today,

\begin{abstract}
The Organic Materials Database (OMDB) is an open database hosting about 22,000 electronic band structures, density of states and other properties for stable and previously synthesized 3-dimensional organic crystals. The web interface of the OMDB offers various search tools for the identification of novel functional materials such as band structure pattern matching and density of states similarity search. In this work the OMDB is extended to include magnetic excitation properties. For inelastic neutron scattering we focus on the dynamical structure factor $S(\mathbf{q},\omega)$ which contains information on the excitation modes of the material. We introduce a new dataset containing atomic magnetic moments and Heisenberg exchange parameters for which we calculate the spin wave spectra and dynamic structure factor with linear spin wave theory and atomistic spin dynamics. We thus develop the materials informatics tools to identify novel functional organic and metalorganic magnets.
\end{abstract}

%\pacs{Valid PACS appear here}


\maketitle

\section{\label{sec:intro}Introduction}
Magnetism and magnetically ordered materials have played a crucial role in the development of the technology used in our every-day life. The identification of novel materials with desired magnetic target properties as well as the investigation of coupling mechanisms, the resulting order and its excitations are therefore of great importance. While basic concepts are usually explored in materials with feasible complexity, materials with complex unit cells and dominant interaction effects quite often exhibit the more desirable properties with respect to technological applications. In particular metal organic frameworks and organic molecular crystals exhibit promising structures for electron-spin-based devices. Magnetic organics have attracted attention with respect to spintronic devices \cite{Spintronics1, Spintronics2, Spintronics3} and magnon spintronics \cite{Liu2018}, multiferroic phases \cite{Stroppa2011,Walker2017}, molecular qubits \cite{qubit1,qubit2}, and spin-liquid physics \cite{Pustogow2018,spinliquid2,spinliquid3}. Local magnetic moments in organic materials can arise due to transition metal and rare earth ions embedded in the molecules or due to local unsaturized bonds as they occur in stable organic radicals \cite{garcia1998magnetic,Miller2011,Miller2014}. Some organics establish stable magnetic order of various kind up to room temperature \cite{manriquez1991room,jain2007high,zaidi2004room}. For device engineering and technological applications, an important advantage of organic materials over other functional materials such as transition metal oxides, is that organics can be synthesized with relative low cost and large-scale production methods.

The vast increase of experimental and theoretical data obtained over the past century has opened a novel approach to materials research based on computer science methods and the construction of materials databases \cite{materials_project,Borysov2017,draxl2018nomad,ortiz2009data,pubchemqc,drugbank,schomburg2002brenda,Herper2017}. Such databases were successfully applied in mining for functional materials \cite{ortiz2009data,klintenberg2013possible,klintenberg2014computational,yang2012search,glawe2016optimal,Geilhufe2017a,Geilhufe2017,geilhufe2018materials} and as training data for machine learning algorithms predicting complex materials properties \cite{De2016,Schtt2018,Xie2018,olsthoorn2019}, bypassing computationally demanding \emph{ab initio} calculations. In this article we report about the implementation of a novel dataset for organic magnets into the organic materials database OMDB \cite{Borysov2017,omdb}, available at \url{https://omdb.mathub.io}. The data was calculated by means of \emph{ab initio} calculations and comprises information about local magnetic moments, magnetic exchange coupling, expected magnetic ground state, as well as spin-wave excitation spectra and the dynamical structure factor $S(\mathbf{q},\omega)$ for hundreds of previously synthesized organic molecular crystals and metal organic frameworks. This data is embedded into the existing framework of the OMDB and can be explored using interactive statistics and non-trivial search tools such as pattern matching \cite{Borysov2018,Geilhufe2018a}.
 
The implementation of such a database is closely related to the recent enhancement of neutron flux \cite{ess,sns,jparc} and detector technology, allowing for inelastic neutron scattering experiments to be performed with higher signal to noise ratio, and higher energy resolution \cite{Birk2014,Groitl2016}. For example, for inelastic neutron scattering experiments, the central entity is the dynamical structure factor $S(\mathbf{q},\omega)$, which contains information on the excitation modes of the material. The vast majority of inelastic neutron scattering (INS) measurements of magnetic materials are analyzed by means of fitting the experimental dynamical structure factor $S(\mathbf{q},\omega)$ to the dispersion and structure factor of a spin wave Hamiltonian as provided in our dataset.

\begin{figure}
\includegraphics[width=0.45\textwidth]{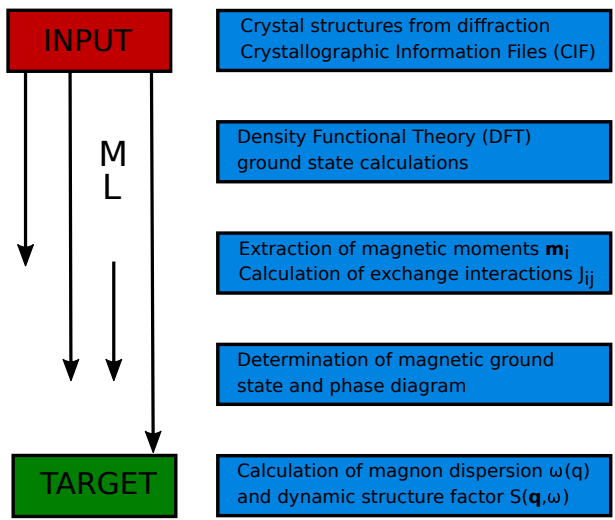}
\caption{\label{fig:workflow}
The workflow for high throughput calculation of magnetic ground states and spin wave spectra. The blue boxes indicate the steps we use for the results presented in this article. The black arrows on the left hand side indicate steps which can be performed with machine learning (ML) techniques, to replace computationally expensive DFT and ASD simulations, ultimately enabling for prediction of magnon spectra (green TARGET box) directly from the crystal structure (red INPUT box).
}
\end{figure}

The remaining of the article is organized as follows. In Section~\ref{sec:methods} we present the models and approximations used, as well as the scheme developed for high throughput calculation of magnetic properties and excitation spectra. Results are presented in Sec. \ref{sec:results}, followed by a conclusion in Sec. \ref{sec:conclusions} and outlook in Sec. \ref{sec:outlook}.

\section{\label{sec:methods}Models and Methods}
A state-of-the-art approach for \emph{ab initio} investigations of spin dynamics in magnetic solids is to solve the full electronic dynamics with time-dependent density functional theory (TD-DFT) and calculating the energies and lifetimes of the spin waves without resorting to the adiabatic approximation \cite{Buczek2009,Krieger2015}. However, both the real time and the linear response variants of TD-DFT are computationally demanding and currently not feasible for organic materials with large chemical unit cells containing in average $\approx 80$ atoms for the materials stored within the OMDB. The calculation of the coupling parameters for magnetic Hamiltonians is nowadays supported by various DFT softwares, such as RSPt~\cite{Wills2010fes}, SPR-KKR~\cite{Ebert2011}, KKRnano \cite{Thiess2012}, and HUTSEPOT \cite{luders2001ab,fischer2009}. This opens the path for a more tractable approach in context of the high-throughput computations performed by us, allowing for an \emph{ab initio} modeling of spin wave excitation spectra in form of a two-step approach: first, the coupling parameters of an effective magnetic Hamiltonian are calculated from DFT calculations; second, analytical or numerical spin wave theory calculations are performed to obtain the spin wave spectra. Furthermore, thermodynamic properties of the spin Hamiltonian can be investigated with Monte Carlo and atomistic spin dynamics (ASD) simulations \cite{Eriksson2017asd}.

The description of our methodology is contained in a number of sections. The magnetic model Hamiltonian and details about the density functional theory calculations go into Sec. \ref{sec:MagHam}, followed by a brief description of linear spin wave theory for collinear magnets in Sec. \ref{sec:ams}, and the atomistic spin dynamics method in Sec. \ref{sec:ASD}. The high throughput workflow is described in Section \ref{sec:highput} and is shown schematically in Fig. \ref{fig:workflow}. The pattern matching method for magnon dispersions is described in \ref{sec:matcher}.

\subsection{\label{sec:MagHam}Magnetic Hamiltonian}

For a magnetic solid with $N_A$ atoms in the crystallographic unit cell and a total of $N_c$ cells, we use the notation $\mathbf{R}_{i\mu}=\mathbf{R}_{i}+\mathbf{r}_{\mu}$ to specify atomic positions, where $\mathbf{r}_{\mu}$ is the position of the basis atom $\mu$ in the unit cell, and $\mathbf{R}_{i}$ is the position of the crystallographic unit cell $i$. The magnetic model is a Heisenberg Hamiltonian
\begin{equation}
\label{eq:HeisHam}
\mathcal{H} = -\frac{1}{2} \frac{1}{N_c N_A} \sum_{i\mu}^{N_c N_A} \sum_{j\nu}^{N_c N_A} J_{i\mu j\nu} \hat{\mathbf{e}}_{i\mu} \cdot \hat{\mathbf{e}}_{j\nu},
\end{equation}
formulated in terms of unit vectors $\hat{\mathbf{e}}_{i\nu}$ for the directions of the local magnetic moment at site $i\mu$. Note the sign convention with a leading minus sign for the first term, and that the double summation
$\sum_{i\mu}^{N_cN_A} \sum_{j\nu}^{N_cN_A}$ count each bond twice. The magnetic moment of an atom is $\boldsymbol{\mu}_{i\mu}=\mu_{\rm B}m_{\mu}\hat{\mathbf{e}}_{i\mu}$, where $m_{\mu}$ is the size of the magnetic moment of atomic type $\mu$ in terms of multiples of Bohr magnetons $\mu_{\rm B}$.

In order to formulate spin Hamiltonians for magnetic organic materials we have used the full-potential linear muffin-tin method (FP-LMTO) software RSPt~\cite{Wills2010fes} to calculate Heisenberg exchange parameters. The full-potential basis set allows for accurate electronic calculations irrespective of the geometry of the crystal structure. The latter is important for the sparse unit cells typically present in organics. The accuracy and reproducibility of DFT results when using different basis sets have recently been investigated in Ref. \cite{Lejaeghere2016}. The implementation of FP-LMTO in RSPt is described in detail in Ref. \cite{Wills2010fes}. Furthermore, in RSPt Heisenberg exchange interactions can be calculated from Green functions for a reference spin structure, \emph{e.g.} a ferromagnetic or collinear antiferromagnetic ordering, by means of the Liechtenstein-Katsnelson-Antropov-Gubanov (LKAG) formalism \cite{Liechtenstein1987}, without the need to set up supercells with different spin configurations. Details on the implementation of the LKAG formula in RSPt can be found in Ref. \cite{Kvashnin2015}.

The calculations were made with the local density approximation (LDA) for the exchange-correlation potential and using a 6 6 6 k-point mesh. The charge density and the potential inside the muffin-tin spheres were represented using an angular momentum decomposition up to $l_{max}$ = 8. One energy set was used for the valence electrons. For the description of the states in the interstitial region three kinetic energy tails were used: -0.3, -2.3, and -1.5 Ryd.

\subsection{\label{sec:ams}Adiabatic magnon theory}
The excitation spectra of a spin Hamiltonian can be obtained by various analytical and semi-analytical methods. The traditional approach relies on the introduction of Holstein-Primakoff (HP) operators for the bosonic operators, and expanding the spin Hamiltonian to quadratic or higher order in these HP operators. The expansion to quadratic order leads to linear spin wave theory which has established itself as the main formalism to model inelastic neutron scattering (INS) data of magnetic materials. This technique and aspects of how it can be implemented in software is described for the general noncollinear case of multisublattice magnets in Refs. \cite{Haraldsen2010,Toth2015,Fishman2018sta}. In the following we briefly review the main steps of the so called adiabatic magnon theory, a formalism that naturally connects to the frozen magnon technique for obtaining spin wave energies from \emph{ab initio} calculations. More details can be found in Ref. \cite{Kubler2000toi}. The spatial Fourier transform of the exchange interaction is defined according to
\begin{eqnarray}
\label{eq:JQ}
J_{\mu \nu}(\mathbf{q})=\sum_{j\ne0}^{N_c}
J_{0\mu j\nu} e^{ i\mathbf{q} \cdot \mathbf{R}_{0\mu j\nu} },
\end{eqnarray}
where $\mathbf{R}_{0\mu j\nu}=\mathbf{R}_{0}+\mathbf{r}_{\mu}-\mathbf{R}_{j}-\mathbf{r}_{\nu}$, and $i=0$ is the central unit cell. After linearizing the equation of motion for the Heisenberg Hamiltonian in Eq. \eqref{eq:HeisHam} for small cone angle excitation relative to the magnetization quantization direction, here chosen to be along $\hat{\mathbf{z}}$, we can introduce the quantity 
\begin{eqnarray}
\label{eq:JQ2}
\tilde{J}_{\mu \nu}(\mathbf{q}) = -\frac{e_{\nu}^z}{m_{\mu}} J_{\mu \nu}(\mathbf{q})
+ \delta_{\mu \nu} \frac{1}{m_{\mu}}\sum_{\lambda}^{N_A} e_{\lambda}^z J_{\mu \lambda}(\mathbf{0})
\end{eqnarray}
in which $e_{\mu}^z=\{1,-1\}$ specify the collinear (anti)ferromagnetic groundstate of the system. The spin-wave dispersion as a function of wave vector $\mathbf{q}$ can then be obtained by diagonalizing $\tilde{J}_{\mu \nu}(\mathbf{q})$.

\begin{figure}
\includegraphics[width=0.45\textwidth]{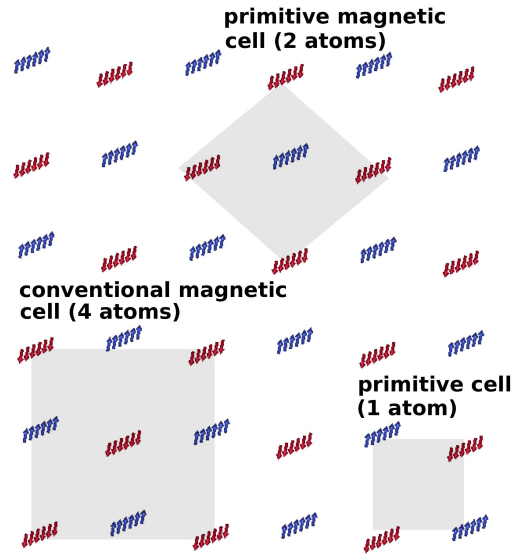}
\caption{\label{fig:MagCellReduction}
Reduction to the primitive magnetic cell for the collinear antiferromagnetic material C$_{20}$H$_{10}$CoN$_{12}$ (OMDB-ID 11913).
%(OMDB-ID 11913, COD-ID 2014058). 
The primitive chemical cell has only one magnetic site. In the ground state search with ASD simulation a conventional magnetic cell with four magnetic sites is found, a cell which can be reduced to a primitive magnetic cell with two magnetic sites.}
\end{figure}

\subsection{\label{sec:ASD}Atomistic spin dynamics}
In common with spin wave theory a core element in the atomistic spin dynamics method \cite{Skubic2008,Eriksson2017asd} is that it is possible to parametrize the energies and dynamics of the magnetic system to a magnetic Hamiltonian formulated in terms of local spins (or magnetic moments) as in Eqn. \eqref{eq:HeisHam}. The other core element is the stochastic Landau-Lifshitz-Gilbert equation (SLLG), 
\begin{eqnarray}\label{eq:sllg1}
\frac{d\bfm_i}{dt} &=&-\frac{\gamma}{(1+{\alpha}^2)} \bfm_i \times [\bfbb_{i}+\bfbf_{i}]\\
&& -\frac{\gamma}{(1+{\alpha}^2)} \frac{\alpha}{m_i} \bfm_i \times \{\bfm_i \times [\bfbb_{i}+\bfbf_{i}]\},\nonumber
\end{eqnarray}
which is a semi-classical equation used to model the motion of the atomic magnetic moments at zero or finite temperature. The first term of the equation is the precessional motion, while the second term describes the damping motion. $\bfm_i=\bfm_i(t)$ is the magnetic moment and experiences an effective magnetic field $\bfbb_i=\bfbb_i(t)$, calculated from the Hamiltonian as $\bfbb_i=-\frac{\partial \mathcal{H}}{\partial \bfm_i}$. $\gamma$ is the gyromagnetic ratio, $\alpha$ is a scalar (isotropic) Gilbert damping constant. Temperature is included in the form of Langevin dynamics with the power of the stochastic magnetic field $\bfbf_i=\bfbf_i(t)$ related to the damping constant $\alpha$ via a fluctuation-dissipation relation \cite{Eriksson2017asd}. For the ASD simulations we use the UppASD software \cite{UppASD}.

In numerical simulations of the stochastic Landau-Lifshitz-Gilbert equation there is no need for a linearization of the equations of motion wherefore the full dynamics of a magnetic system described by a bilinear-in-spin-operators Hamiltonian such as Eqn. \eqref{eq:HeisHam} is retained \cite{Hellsvik2009,Eriksson2017asd}. Furthermore, also the full dynamics of a magnetic Hamiltonian which contains higher order coupling such as biquadratic exchange can be studied without the need for mean-field treatment of these couplings, see \emph{e.g.} Ref. \cite{Hellsvik2016}. Spatial and temporal fluctuations are sampled using a time-dependent correlation function
\begin{equation} 
\label{eq:cqt}
C^{\alpha \beta}(\mathbf{r},t)=\frac{1}{N}\sum_{\substack{i,j ~ \text{where} \\\ ~\mathbf{r}_i-\mathbf{r}_j=\mathbf{r}}} \langle m_i^\alpha(t) m_j^\beta(0) \rangle,
\end{equation}
where $\alpha$ and $\beta$ are the Cartesian components of the magnetic moments. Fourier transforming $C(\mathbf{r},t)$ over space and time yields the dynamic structure factor
\begin{equation} 
\label{eq:sqw}
S^{\alpha \beta}(\mathbf{q},\omega)=\frac{1}{\sqrt{2\pi}N}\sum_{\mathbf{r}} e^{i\mathbf{q}\cdot\mathbf{r}} \int_{-\infty}^\infty e^{i\omega t}C^{\alpha \beta}(\mathbf{r},t) \,dt,
\end{equation}
which can be related to the scattering intensity measured in inelastic neutron or electron scattering on magnetic materials. Only the magnetic excitations perpendicular to the scattering wave vector contribute to the scattering intensity \cite{White2007qto,Fishman2018sta}
\begin{equation} 
\label{eq:sqwintensity}
S(\mathbf{q},\omega)=\sum_{\alpha,\beta}\left( \delta_{\alpha\beta} - \frac{q_{\alpha}q_{\beta}}{q^2}\right)S^{\alpha \beta}(\mathbf{q},\omega).
\end{equation}
For the measurement of the dynamic structure factor ASD simulations were run at $T=1$ K using a time step $dt=5\cdot10^{-16}$~s and a small Gilbert damping $\alpha=0.0001$.
The correlation function in Eq. \eqref{eq:cqt} was measured using a sampling step of $t_{\rm samp}=5\cdot10^{-15}$~s, over a sampling window of $t_{\rm win}=5\cdot10^{-11}$~s. The corresponding frequency range for the dynamic structure factors in Eqs.~\eqref{eq:sqw} and \eqref{eq:sqwintensity} is $\omega/(2\pi)=[0.02, 0.04,\ldots,200]$~THz (0.0827 meV to 827 meV). In the figures contained in Section \ref{sec:results}, the scattering intensity is shown relative to the correlation $S_0(\mathbf{q},\omega=0)$ for the fully ordered system.

\subsection{\label{sec:highput}High throughput calculation}
In this section is described the methodology that we have developed for high throughput calculation of magnetic ground states and spin wave spectra. The main steps are shown in Figure \ref{fig:workflow}.

Input files for the RSPt ground state calculations are prepared from the crystallographic information files (CIF) using the cif2cell program \cite{Bjorkman2011}. The calculations are performed for a ferromagnetic ground state configuration, without any a priori consideration of what the magnetic ground state could be. In order to classify the atoms as magnetic or non-magnetic, a simple criteria is used, namely that the spin-polarized charge density integrated over the muffin-sphere of the atomic sites is larger than 0.1 $\mu_B$. For these atomic sites Heisenberg exchange interactions are calculated using the approach described in Sec. \ref{sec:MagHam}.

The linear spin wave calculation require as input not only the parameters of the magnetic Hamiltonian, but also ground state spin configuration. In order to search for the ground state we use an atomistic spin-dynamics quenching scheme for simulation cells with edge length $L$, corresponding to a number $N_{\rm atom}=N_A L^3$ of sites, where $N_A $ is the number of magnetic atoms in the primitive chemical cell. Starting from having all magnetic moments initially in a ferromagnetic configuration, the system evolves in Langevin dynamics simulation at finite temperature. This is followed by a sequence of simulations at zero temperature but with finite Gilbert damping $\alpha$. For zero or static external magnetic field, the deterministic Landau-Lifshitz equation has the property that energy is a non-increasing function of time, and consequently that the energy of the spin Hamiltonian will be minimized \cite{Eriksson2017asd}. For systems with competing magnetic phases at low temperatures, there is a finite probability that the system will get caught in a local minimum so that the true ground state of the magnetic Hamiltonian is not found, an issue that could be of particular concern for systems with highly degenerate (free) energy landscapes such as frustrated magnets and spin glasses \cite{Skubic2009a}. We do not have quenched chemical disorder wherefore a spin glass phase is not possible, furthermore a pure Heisenberg Hamiltonian as in Eq. \ref{eq:HeisHam} cannot stabilize a skyrmion spin texture. For the present data set, we classify the systems as ferromagnetic, collinear antiferromagnetic, noncollinear antiferromagnetic, or paramagnetic. An overview and statistics of the dataset will be presented in Section \ref{sec:results}. Another important distinction is whether the magnetic ordering is commensurate with the primitive chemical cell or not. Knowledge of the shape and size of the primitive magnetic cell, allows for the linear spin wave calculation to be performed for the size of the spin wave Hamiltonian that will give the correct number of magnon dispersion eigenvalues and eigenvectors. In order to obtain from a size $N_A L^3$ simulation cell the smallest possible magnetic cell, we make use of the capability of the Elk software \cite{elk} to reduce from an input crystallographic structure to the smallest primitive cell, considering also the magnetic ordering.

For the DFT ground state calculations a typical value of the computation time was 125 hours on a compute node with 32 cores (4 000 core hours). The calculations of the magnetic ground states and the dynamic structure factor with atomistic spin dynamics are more than one order of magnitude faster than the DFT calculations. The sampling of the dynamic structure factor takes of the order of 10 hours using 8 cores (80 core hours). The calculation of the magnon dispersion for collinear spin configurations is quasi instantaneous, as it only involves the diagonalization of small matrices.

\subsection{\label{sec:matcher}Magnon matcher}
Dispersion relations for magnons are commonly calculated along lines in reciprocal space between high symmetry points in the Brillouin zone. We have chosen to use for each material the same paths for magnons dispersions as has been used for the creation of the OMDB electronic band structure data set \cite{Borysov2017,omdb}.

Similar to the previously described pattern matching for electronic band structures \cite{Borysov2018}, we have implemented a pattern matching algorithm for magnon bands. The pattern matching algorithm is based on a moving window approach which is best described for a pattern containing two bands (it can be extended to an arbitrary number of bands). A user can draw a two-band pattern in the OMDB user interface \url{https://omdb.mathub.io/search/pattern_magnon}. The pattern is vectorized and distributed in two vectors $\mathbf{v}_{\mathrm{upper}}$ and $\mathbf{v}_{\mathrm{lower}}$ which are concatenated into a query vector $\mathbf{v}_{\mathrm{query}}=\mathbf{v}_{\mathrm{upper}},\mathbf{v}_{\mathrm{lower}})$.

For each calculated magnon spectrum we consider pairs of magnon branches and choose a window of a user specified width. The two bands within the window are vectorized in a similar way as the query vector and compared to the query vector in terms of a vector distance, here the cosine distance. The window is moved by a specified stride and the resulting vectors are compared again. Finally, from all magnon spectra and all possible windows the ones which have a suitable vector similarity to the initial pattern are selected. This algorithm is implemented on the OMDB web interface, where a ball tree nearest neighbour search is used to accelerate the comparison between initial pattern and respective windows within the magnon spectra. More details on the method can be found in Ref. 38.

\section{\label{sec:results}Results}
The main dataset of the OMDB constitute of spin-polarized electron band structures and density of states \cite{Borysov2017,omdb}. 
The first step in the production of the current set of magnetic Hamiltonians and excitation spectra was to choose a subset of the magnetic materials on the OMDB. For these materials we have performed modeling using the high throughput methodology described in Sec. \ref{sec:highput}. We have obtained the magnetic excitation spectra for 118 materials, containing one or more sites with a magnetic moment of at least 0.1 $\mu_{\rm B}$ according to the RSPt LDA calculations. The majority of the materials in this magnetic excitations dataset have noncollinear ground states. Rather remarkably, none of our currently considered materials displays a ferromagnetic groundstate. The ASD quenching simulations to obtain magnetic ground states used simulation supercells with edge length $L=6$, and the ASD simulations for sampling $S(\mathbf{q},\omega)$ at finite temperature used supercells with $L=38$.

We will in the following present our magnetic excitations dataset (OMDB-SW1), closely connecting to how the results are presented on the Organic Materials Database. We refer to the materials using their OMDB-ID numbers.
%We refer to the materials using their ID numbers (COD-ID) within the crystallographic open data base (COD) \cite{Graulis2009,COD}, and the OMDB-ID numbers.
Our first example is $1,1',2,2',4,4'$-hexaisopropylnickelocene C$_{28}$H$_{46}$Ni (OMDB-ID 855) 
%(OMDB-ID 855, COD-ID 4064866) 
which has very low symmetry and crystallizes in space group $P$ -1 (space group number 2). The compound is formed from isopropyl groups and of cyclopentadienyl rings that sandwich divalent Ni atoms \cite{Weismann2011}. The primitive nonmagnetic cell has 75 atoms. The Ni atom has Wyckoff position 1a which has inversion symmetry. For this site we obtained a magnetic moment $\mu_{\rm Ni}=0.84 \mu_{\rm B}$. Heisenberg interactions were calculated between the Ni sites for distances up to a maximum of three multiples of the lattice constant $a$. All the leading exchange couplings are antiferromagnetic. The shortest Ni to Ni distance of 8.75 {\AA} is found along the crystallographic $a$ axis with the Ni spins coupled antiferromagnetically, $J_{8.75}=-0.012$ meV. The next nearest neighbour coordination shell of Ni atoms is along the $b$ axis with distance 9.08 {\AA} and the vanishingly small value coupling $J_{9.08}=-0.002$ meV. The strongest exchange coupling $J_{9.16}=-0.056$ meV is found for the third nearest neighbor coordination shell in the $ab$-plane. Three dimensional coordination of exchange between Ni atoms is achieved when considering the coupling $J_{9.26}=-0.012$ meV along the $c$ axis. On the OMDB \cite{omdb}, the exchange couplings are listed, ordered to show the strongest couplings first, in a table in the lower left part of the magnetic properties panel and are also displayed in the JSmol viewer. A listing of the magnetic moments can be found in the lower right part of the panel.

\begin{figure}
\includegraphics[width=0.40\textwidth]{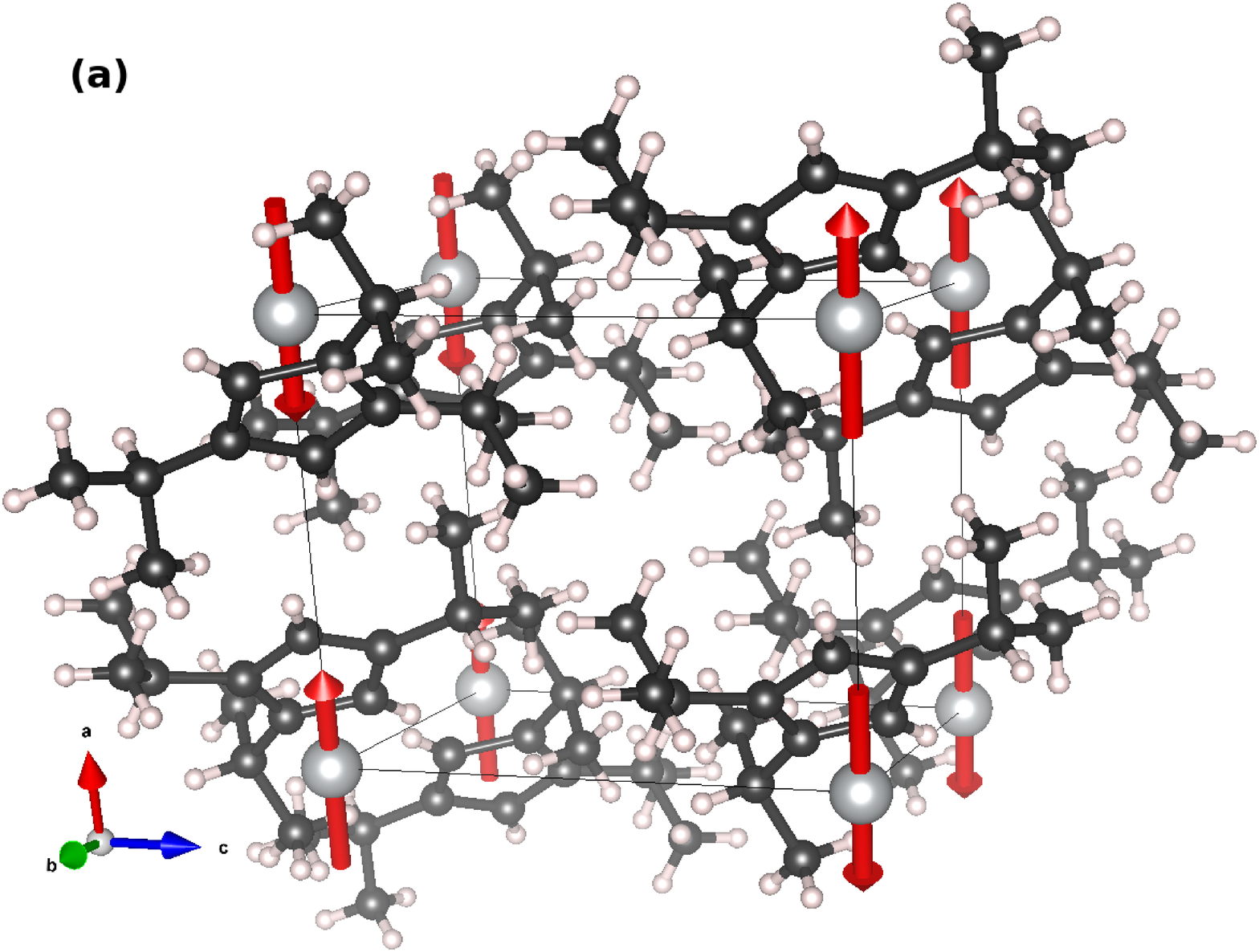}
\includegraphics[width=0.50\textwidth]{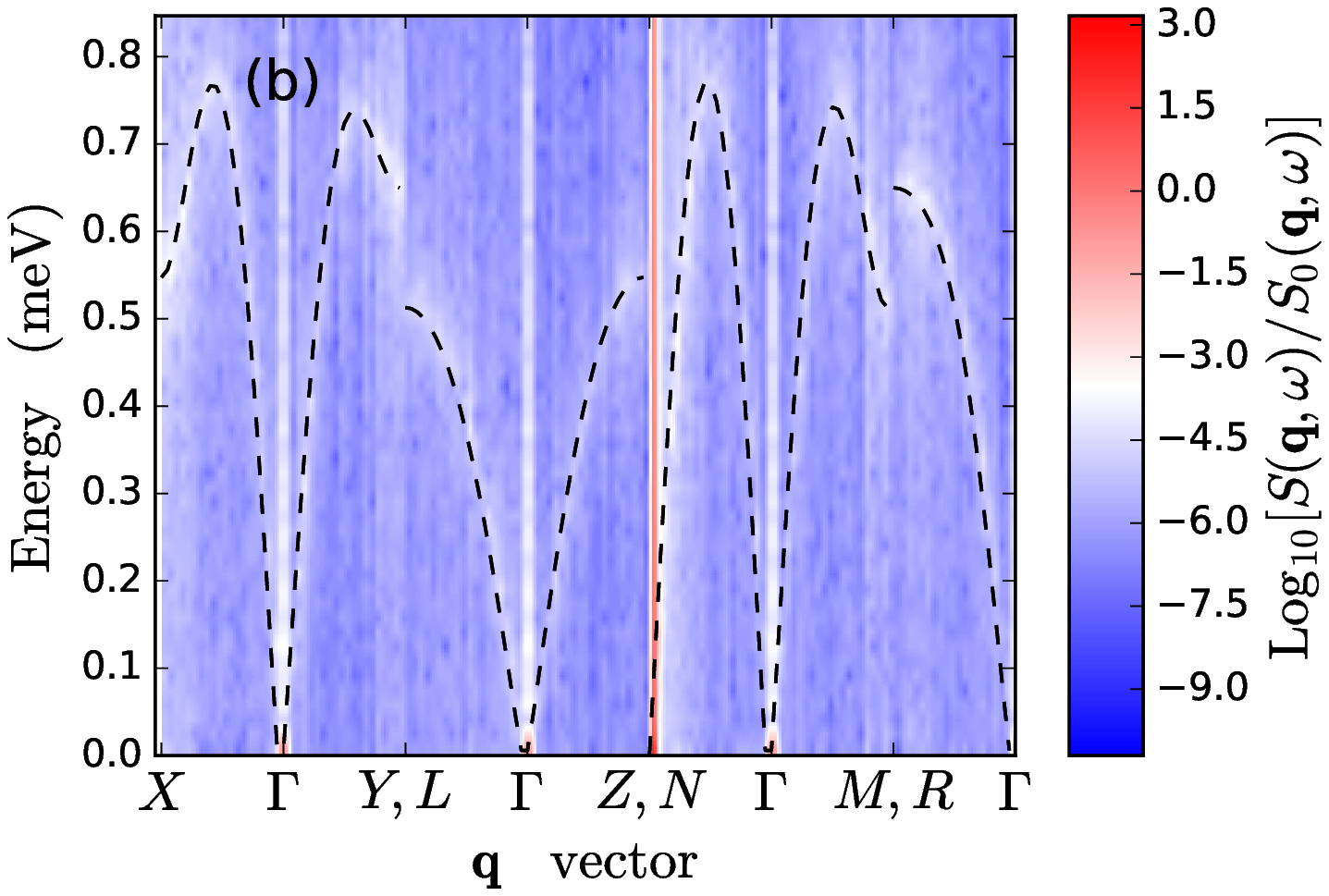}
\caption{\label{fig:4064866}Magnetic properties of the collinear antiferromagnetic hexaiso-propylnickelocene
C$_{14}$H$_{23}$Ni$_{0.5}$ (OMDB-ID 855) 
%(OMDB-ID 855, COD-ID 4064866) 
(a) The crystallographic unit cell with its magnetic sites and ground state spin configuration shown as red arrows. (b) The $T=0$ K magnon dispersion (black dashed line) and the $T=1$ K dynamical structure factor (colorplot).
}
\end{figure}

The outcome of the ASD quenching calculation was a collinear antiferromagnetic spin configuration that is shown in Fig. \ref{fig:4064866}(a). With regard to the original lattice vectors for the primitive chemical cell, a $2\times 1\times 2$ supercell of the primitive cell can accommodate the antiferromagnetic ordering. Calculating the spin wave spectra for this conventional magnetic cell, two doubly degenerate magnon bands come out (not shown). The conventional magnetic cell can be reduced to a two-atom magnetic cell as shown schematically in Fig. \ref{fig:MagCellReduction}. For this primitive magnetic cell one doubly degenerate magnon band comes out. Figure \ref{fig:4064866}(b) displays the $T=0$ K magnon dispersion (black dashed line) along a high symmetry path in the Brillouin zone as well as the corresponding $T=1$ K dynamical structure factor intensity (colorplot), sampled using Eqs. \eqref{eq:cqt}-\eqref{eq:sqwintensity}.

\begin{figure}
\includegraphics[width=0.40\textwidth]{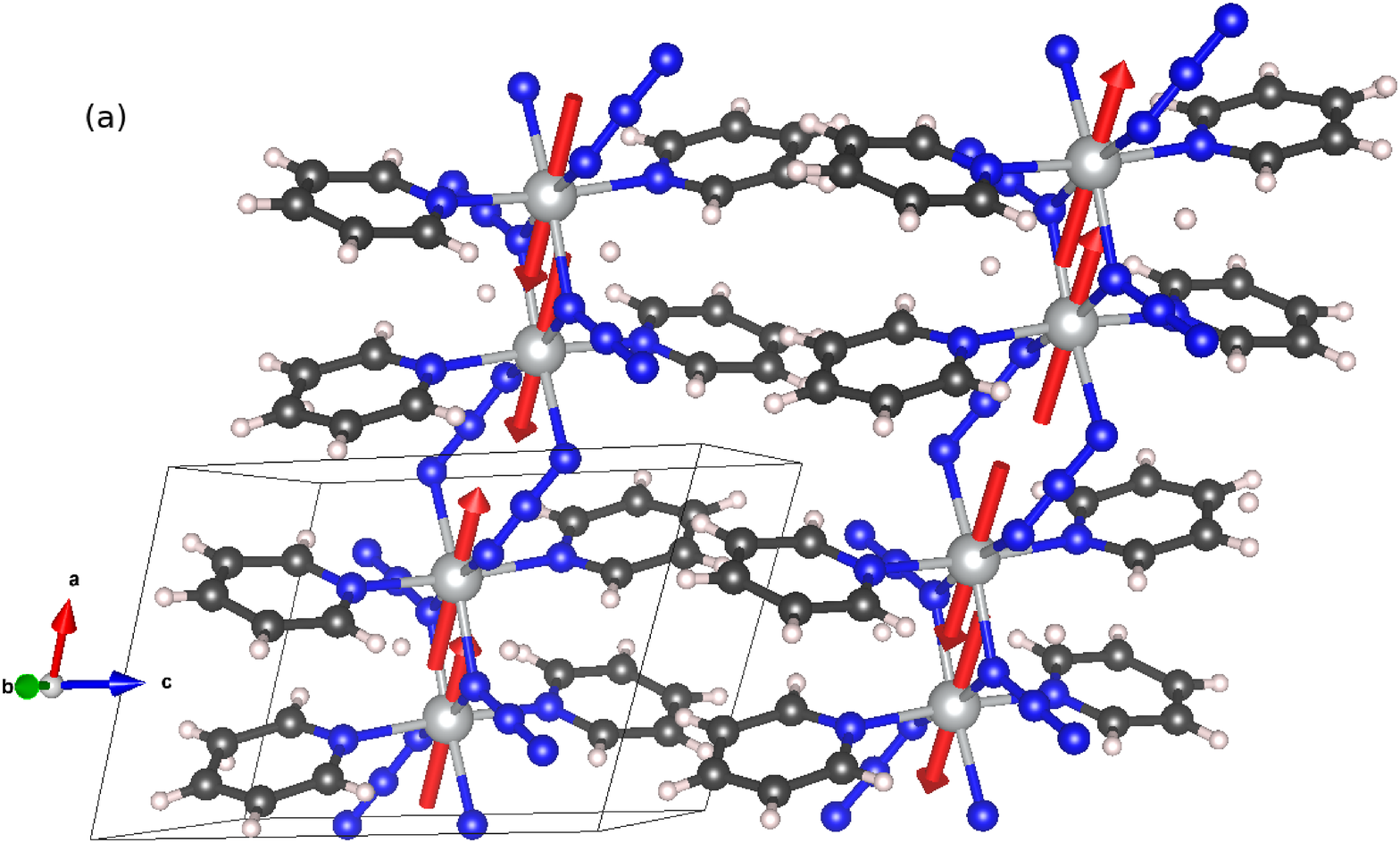}
\vfill
\includegraphics[width=0.50\textwidth]{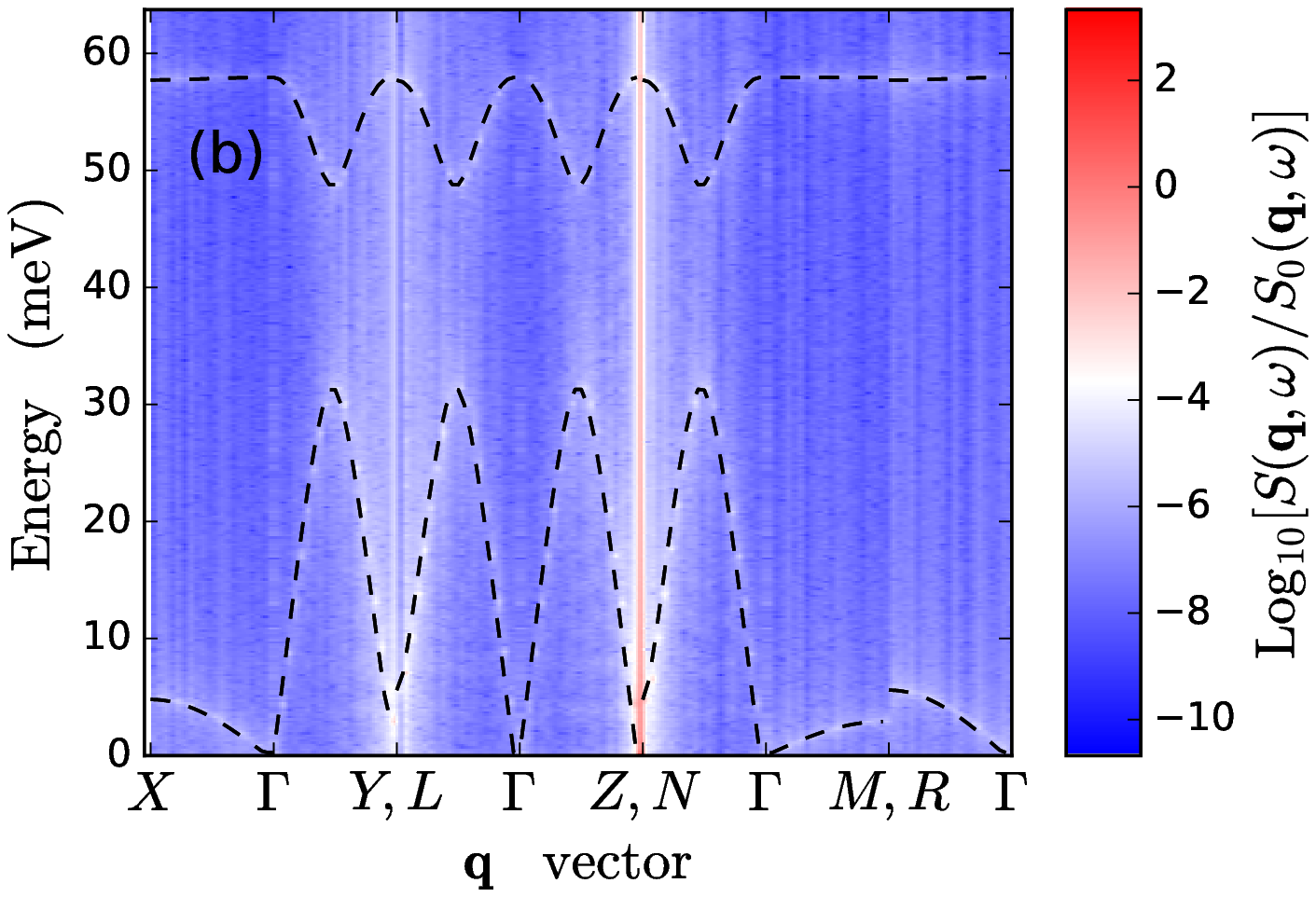}
\caption{\label{fig:2216760}Magnetic properties of the collinear antiferromagnetic azido-Ni compound C$_{10}$H$_{10}$N$_{8}$Ni (OMDB-ID 19963).
%(OMDB-ID 19963, COD-ID 2216760).
(a) The crystallographic unit cell with its magnetic sites and ground state spin configuration shown as red arrows. (b) The $T=0$ K magnon dispersion (black dashed line) and the $T=1$ K dynamical structure factor (colorplot).}
\end{figure}

Also the azido-Ni compound C$_{10}$H$_{10}$N$_{8}$Ni \cite{Song2007} (OMDB-ID 19963) 
%(OMDB-ID 19963, COD-ID 2216760) 
has a structure that belongs to space group $P$ -1 (space group number 2). The primitive nonmagnetic cell has 58 atoms, corresponding to two formula units. For the Ni site we obtain a magnetic moment $\mu_{\rm Ni}=1.05\mu_{\rm B}$. The magnetic ground state is collinear, and can be accommodated in a $2\times 1\times 2$ supercell of the primitive cell. Given that the primitive nonmagnetic cell has two Ni sites, the supercell contains a total of eight magnetic sites. Similarily as for C$_{20}$H$_{10}$CoN$_{12}$, the supercell can be reduced to a primitive magnetic cell, this time with four atoms. The linear spin wave calculations yield two doubly degenerate magnon bands, one pair being an acoustic mode, and the other pair being an optical mode with finite values of the dispersion at the zone center. Our results for C$_{10}$H$_{10}$N$_{8}$Ni are displayed in Fig. \ref{fig:2216760}. The color plot for the dynamic structure factor gives an indication on the relative occupancy of higher and lower energy magnons at the given temperature of $T=1$ K \footnote{The input files for atomistic spin dynamics simulations are posted on the OMDB materials pages, enabling for straightforward simulation and sampling of the dynamic structure factor at other temperatures using the UppASD software \cite{UppASD}}.

\begin{figure}
\includegraphics[width=0.50\textwidth]{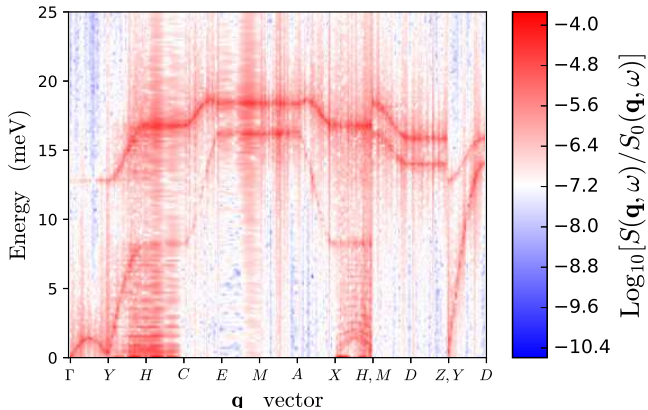}
\caption{\label{fig:1515486}The $T=1$ K dynamical structure factor (colorplot) for manganese dicarboxylate C$_2$H$_3$MnO$_3$ (OMDB-ID 11283). 
%(OMDB-ID 11283, COD-ID 1515486).
Unlike for the collinear magnets shown in Figs. \ref{fig:4064866} and \ref{fig:2216760}, we here do not include any $T=0$ K magnon dispersion, as the current workflow does not support high throughput calculation of linear spin wave spectra for noncollinear spin textures. 
}
\end{figure}

As a third example, we will discuss the noncollinear antiferromagnet material manganese dicarboxylate C$_2$H$_3$MnO$_3$ \cite{Saines2011} (OMDB-ID 11283),
%(OMDB-ID 11283, COD-ID 1515486),
with results displayed in Fig. \ref{fig:1515486}. The material crystallizes in the non-symmorphic space group $P$ 1 $21/n$ 1 (space group number 14). The primitive chemical cell contains 36 atoms, corresponding to four formula units. A magnetic moment is carried by the Mn sites for which our DFT calculations gave a magnetic moment $\mu_{\rm Mn}=2.20\mu_{\rm B}$. The outcome of the ASD quenching simulation was a canted antiferromagnetic phase. Our current implementation of linear spin wave theory does not support fully automated calculations of the spin wave spectra for systems with noncollinear ground states. In contrast, the ASD simulation method allows for semiclassical sampling of the dynamic structure factor for magnetic systems with any spin ordering, including also the paramagnetic regime where long range ordering is absent. Figure \ref{fig:1515486} shows the dynamical structure factor at $T=1$ K for C$_2$H$_3$MnO$_3$, revealing two main bands of dispersion.

\begin{figure}
\includegraphics[width=0.50\textwidth]{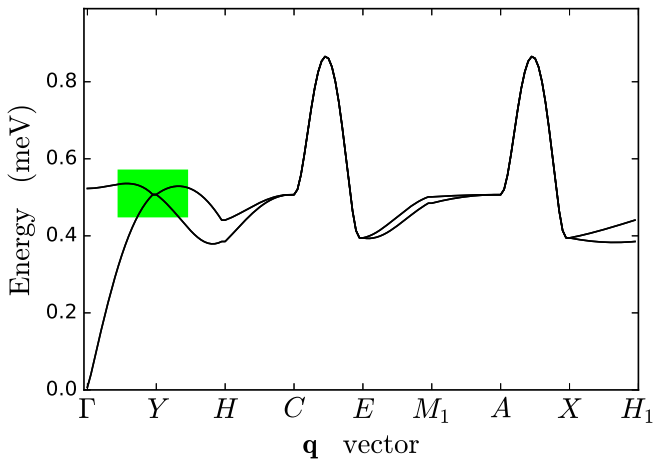}
\caption{\label{fig:exampleMagnonMatcher} Example of use of the magnon matcher for the identification of Dirac nodes of the magnon bands. The black curves show the spin wave dispersion for the space group $P$ 1 $21/n$ 1 (space group number 14) collinear antiferromagnet C$_6$H$_{14}$NiO$_8$ (OMDB-ID 21042), with the green rectangle indicating a linear crossing of magnon bands at the high symmetry point $Y$.  
%(OMDB-ID 21042, COD-ID 4325324). 
}
\end{figure}

Figure \ref{fig:exampleMagnonMatcher} shows the application of the magnon matcher for the example of identifying Dirac nodes within magnon spectra. Using as a query Dirac-type crossing of linear bands, the pattern matching algorithm here identified such a crossing along a line in reciprocal space at the high symmetry point Y, for the space group 
$P$ 1 $21/n$ 1 (space group number 14) collinear antiferromagnet C$_6$H$_{14}$NiO$_8$ (OMDB-ID 21042).
%(OMDB-ID 21042, COD-ID 4325324).
The antiferromagnetic ground state breaks the monoclinic symmetry of the chemical unit cell. The corresponding magnetic space group symmetry was determined using FINDSYM \cite{Stokes2005,findsym} and is given by $P_S-1$, having the coset representatives $E=\{1|0\}$, $I=\{-1|0\}$, $T = \{1'|t\}$, $TI =\{-1'|t\}$, with $t=\{0,0,1/2\}$ in units of the real space lattice.
Here, $1'$ and $-1'$ denote the identity $(x\to x,y\to y,z\to z)$ and the inversion $(x\to -x,y\to -y,z\to -z)$ connected with a flip of the magnetization $(m_x\to -m_x, m_y\to -m_y, m_z\to -m_z)$. 
The magnetic unit cell contains four atoms, with two moments pointing in $z$ and two moments in $-z$ direction. Due to the antiferromagnetism, each mode of frequency $\omega$ and momentum $k$, belonging to magnetic sites pointing in $z$-direction, comes with a corresponding mode of opposite frequency and momentum belonging to the magnetic sites pointing in opposite direction. However, both modes are similar in energy leading to two-fold degeneracy of the magnon energy momentum dispersion within the entire Brillouin zone. At the $Y$ point, inversion symmetry is present and eigenstates of the linear spin wave Hamiltonian can be written in a basis where they are eigenvectors of the inversion operation with respective parity eigenvalues $\lambda_i$. As inversion $I$ squares to $1$, i.e., $I^2=1$, the eigenvalues are $\lambda_+ = 1$ and $\lambda_- = -1$. However, at this point the operation $TI$ squares to $-1$, i.e., $TI^2=-1$, leading to a Kramers-like degeneracy of opposite parity modes. In Figure \ref{fig:exampleMagnonMatcher} a green rectangle indicates this linear crossing of magnon bands at the high symmetry point $Y$.

\section{\label{sec:conclusions}Conclusions}
In summary we have developed a method for high-throughput calculation of the magnetic excitation spectra of organometallic materials. To this end we have deployed a multiscale modeling approach in which first the atomic magnetic moments and interaction coupling constants of magnetic Hamiltonians are calculated, followed by ground state determination, and calculation of the spin wave dispersion and the dynamic structure factor. For the present results we have used Heisenberg Hamiltonians with interactions ranging up to three lattice constants. Frustration among the Heisenberg interactions can lead to noncollinear magnetic ground states, and we have among our materials encountered collinear antiferromagnets as well as noncollinear spin orderings. Currently we have used an implementation of linear spin wave theory for collinear magnets, wherefore dispersions were calculated only for the collinear systems, however, we have sampled the dynamic structure factor at $T=1$~K for all materials using atomistic spin dynamics.

The method can be naturally extended to include other interactions such as Dzyaloshinskii-Moriya interaction, Ising interaction, and single-site magnetocrystalline anisotropy energy to the magnetic Hamiltonian, of relevance for the low-energy excitation spectra of quantum organometallic materials. Related to this we expect an even higher fraction of noncollinear ground states, for which the need for calculation of the spin wave dispersions with the more general framework of linear spin wave theory for noncollinear magnets is desired \cite{Toth2015,Fishman2018sta}. Furthermore, recent developments of machine learning techniques for lattice models and spin Hamiltonians, as for instance a profile method for recognition of three-dimensional magnetic structures \cite{Iakovlev2019}, determination of phase transition temperatures by means of self-organizing maps \cite{Shirinyan2019}, and a support vector machines based method for multiclassification of phases \cite{Liu2019b}, will be most useful for identification and classification of competing magnetic phases at finite temperature, and the corresponding phase transition temperatures. 

\section{\label{sec:outlook}Outlook}
A high-throughput database of magnetic properties of organometallic materials opens the path towards establishing general concepts of materials statistics. The search tools that we have developed for electronic properties \cite{Borysov2017, Geilhufe2018a} with the new data for magnetic properties, will enable search for user-specified patterns in the magnon dispersion relations and magnon density of states.\cite{Borysov2017, Geilhufe2018a}. Such tools provide a novel approach towards identifying functional magnetic materials, such as magnon Dirac materials \cite{pershoguba2018dirac,fransson2016magnon,mook2016tunable}, topological magnon insulators \cite{mook2014edge,zhang2013topological,nakata2017magnonic}, and magnon Hall materials \cite{onose2010observation,mook2014magnon}. While there is a huge presence of theoretical concepts of such novel topological magnon phases, only very few materials are known exhibiting these bosonic quantum phases. We see strong indications that the tools presented by us during this article will help us to efficiently explore the realm of organic materials in this context, and note that the very same method can also be used for the calculation of magnon spectra of inorganic magnetic materials. With the database grown to a decent size we furthermore see great opportunities in training machine learning models. We have recently shown that such machine learning tools can be applied to predict basic materials properties for extremely complex organic materials hosting thousands of atoms in the unit cell and with that are outside the realm of materials which can be calculated straightforwardly using density functional theory \cite{olsthoorn2019}. In a similar manner, we see great opportunities in training models towards automatically formulating effective Heisenberg models for arbitrary organic materials, significantly accelerating the accumulation of magnon and $S(\mathbf{q},\omega)$ spectra, as needed, for instance by large-scale experimental facilities \cite{ess,sns,jparc}.

\section{\label{sec:ack}Acknowledgements}
We acknowledge economic support from the Swedish Research Council (VR) through a neutron project grant (BIFROST, Dnr. 2016-06955), and by VILLUM FONDEN via the Centre of Excellence for Dirac Materials (Grant No. 11744). The authors acknowledge computational resources from the Swedish National Infrastructure for Computing (SNIC) at the Center for High Performance Computing (PDC), the High Performance Computing Center North (HPC2N), and the Uppsala Multidisciplinary Center for Advanced Computational Science (UPPMAX).

%\bibliography{library,rmg_library,softwares,neutronsources}

\end{document}